\documentclass[sn-mathphys-num]{sn-jnl}

\usepackage{graphicx}%
\usepackage{multirow}%
\usepackage{amsmath,amssymb,amsfonts}%
\usepackage{amsthm}%
\usepackage{mathrsfs}%
\usepackage[title]{appendix}%
\usepackage{xcolor}%
\usepackage{textcomp}%
\usepackage{manyfoot}%
\usepackage{booktabs}%
\usepackage[utf8]{inputenc}
\usepackage{algorithm}%
\usepackage{algorithmicx}%
\usepackage{algpseudocode}%
\usepackage{listings}%
\usepackage{setspace}
\usepackage{float}          
\usepackage{booktabs}       
\usepackage{tabularx}       
\usepackage[T1]{fontenc}
\usepackage[utf8]{inputenc} 
\usepackage{url}            
\newcolumntype{L}[1]{>{\raggedright\arraybackslash}p{#1}} 
\usepackage{geometry}
\begin{document}
\doublespacing
\title[Article Title]{Concentrated siting of AI data centers drives regional power-system stress under rising global compute demand}


\author*[1]{\fnm{Danbo} \sur{Chen}}\email{chen.12712@buckeyemail.osu.edu}

\author[2]{\fnm{Zijun} \sur{Zhou}}\email{zijunzhou@meta.com}

\author[1]{\fnm{Yongyang} \sur{Cai}}\email{cai.619@osu.edu}

\author[3]{\fnm{Jiahong} \sur{Qin}}\email{vivian.qjh@foxmail.com}

\author[1]{\fnm{Ani} \sur{Katchova}}\email{katchova.1@osu.edu}

\author[3]{\fnm{Lei} \sur{Chen}}\email{xuehaishizi@163.com}

\affil[1]{The Ohio State University, Columbus, OH 43210, USA}

\affil[2]{Meta Platforms Inc., Menlo Park, CA 94025, USA}


\affil[3]{ Zhejiang A\&F University, Hangzhou, Zhejiang 311300, China}

\abstract{The rapid rise of generative artificial intelligence (AI) is driving unprecedented growth in global computational demand, placing increasing pressure on electricity systems. This study introduces an AI–energy coupling framework that combines large language models (LLMs)–based analysis of corporate, policy, and media data with quantitative energy-system modeling to forecast the electricity footprint of AI-driven data centers from 2025 to 2030. 
Results show that the new AI infrastructure is highly concentrated in North America, Western Europe, and the Asia–Pacific, which together account for more than 90\% of projected compute capacity. Aggregate electricity consumption by the six leading firms is projected to increase from roughly 118~TWh in 2024 to between 239~TWh and 295~TWh by 2030, equivalent to about 1\% of global power demand. Regions such as Oregon, Virginia, and Ireland may experience high Power Stress Index (PSI) values exceeding 0.25, indicating local grid vulnerability, whereas diversified systems such as those in Texas and Japan can absorb new loads more effectively. These findings demonstrate that AI infrastructure is evolving from a marginal digital service into a structural component of power-system dynamics, underscoring the need for anticipatory planning that aligns computational growth with renewable expansion and grid resilience.}

\maketitle

\section{Introduction}\label{sec1}

The rapid emergence of generative artificial intelligence (AI) and large-scale data analytics has triggered an unprecedented expansion of computational demand \citep{FT2025_AICapacity,Storey2025}. 
As AI models grow exponentially in size and complexity, their training and inference require vast computing power and data throughput, driving record investment in high-performance data centers and digital infrastructure \citep{CHO2025,Pilz2025_AIPower,OEF2025_AI_Electricity,Chen2025_AIDataCenters,Stojkovic2025}. 
According to Bain \& Company’s \textit{Global Technology Report 2025}, sustaining the computational requirements of AI expansion could generate nearly US\$2~trillion in annual revenues by 2030—equivalent to the combined GDP of the world’s ten largest emerging economies. 
This “AI infrastructure boom” is transforming data centers into the industrial backbone of the digital economy and, increasingly, a major source of electricity demand \citep{Ngata2025,Pengfei2025}.

Electricity use has risen sharply in parallel with the digital transition. 
The International Energy Agency (IEA, 2025) projects that global data-center electricity consumption will more than double—from about 415~TWh in 2024 to roughly 945~TWh by 2030—with the United States and China accounting for almost 80\% of the increase \citep{IEA2025_EnergyAI}. 
AI-specific facilities rely on GPU-based computation, which enables large-scale parallel processing but consumes up to six times more power than conventional racks, elevating both cooling intensity and peak-load requirements. 
These facilities are increasingly concentrated in regions with abundant renewable resources, low electricity prices, and favorable climates, yet such clustering also amplifies local grid stress and transmission constraints. 
As AI campuses scale up from megawatt to gigawatt levels, ensuring a reliable and low-carbon electricity supply has become a strategic challenge for utilities, regulators, and technology developers\citep{MLSYS2022_462211f6,CHEN2025108880}.

This evolution exposes a critical omission in contemporary energy-system analysis. 
Conventional forecasting approaches assume relatively stable relationships between computing intensity, hardware efficiency, and electricity use\citep{9599719,LIU2020272,Siddik_2021}. 
Yet these assumptions break down under generative-AI workloads whose energy demand scales super-linearly with model size and whose deployment patterns are globally distributed but uneven across regions. 
Existing models rarely capture how firm-level siting strategies, spatial clustering, and grid characteristics jointly shape the geography and magnitude of AI-related electricity demand. 
Consequently, the emerging “compute–energy nexus”—the structural coupling between digital infrastructure and electricity systems—remains poorly quantified and insufficiently incorporated into energy planning and grid-reliability frameworks.

To address this gap, we develop an AI–energy coupling analysis framework that integrates semantic inference from large language models (LLMs) with quantitative energy modelling to assess the system-level implications of AI-driven data-center expansion (see Figure 4. in Methods). The LLM component, trained on a multi-source corpus (2015–2025) of corporate reports, policy documents, and media coverage, employs Hugging Face’s all-MiniLM-L6-v2 embeddings and FAISS indexing for semantic retrieval of firm- and region-specific information on investment, siting, and sustainability related to the data centers\citep{zijun2025}. 
These insights inform a compute–energy mapping model that links AI workload intensity to electrical load, a scenario-based forecasting module that projects electricity demand trajectories from 2025 to 2030, and a regional Power Stress Index (PSI) that evaluates grid-level pressure by comparing projected data-center load with generation capacity. Together, these modules provide a unified workflow for quantifying the spatial, temporal, and systemic impacts of AI-infrastructure growth.

By integrating the interpretive capacity of LLMs with quantitative energy-system analytics, this study establishes an evidence-based framework for anticipating and managing the electricity challenges arising from AI-driven digitalization. 
The framework supports the design of proactive capacity-expansion strategies, renewable coordination mechanisms, and locational planning approaches that align the pace of digital infrastructure growth with grid reliability and resilience objectives. 
Beyond its methodological contribution, the analysis highlights that AI data-center expansion has become a structurally significant force shaping the configuration and evolution of global electricity systems. 
Governing this emerging compute–energy nexus will be central to ensuring that the rapid growth of computational capacity advances in tandem with the development of resilient, equitable, and low-carbon power networks\citep{CHEN2025108880}.

\section{Results}\label{sec2}

\subsection{Global clustering and heterogeneity in AI data‐center siting.}\label{subsec2}
\begin{figure}[H]
\centering
\includegraphics[width=0.9\textwidth]{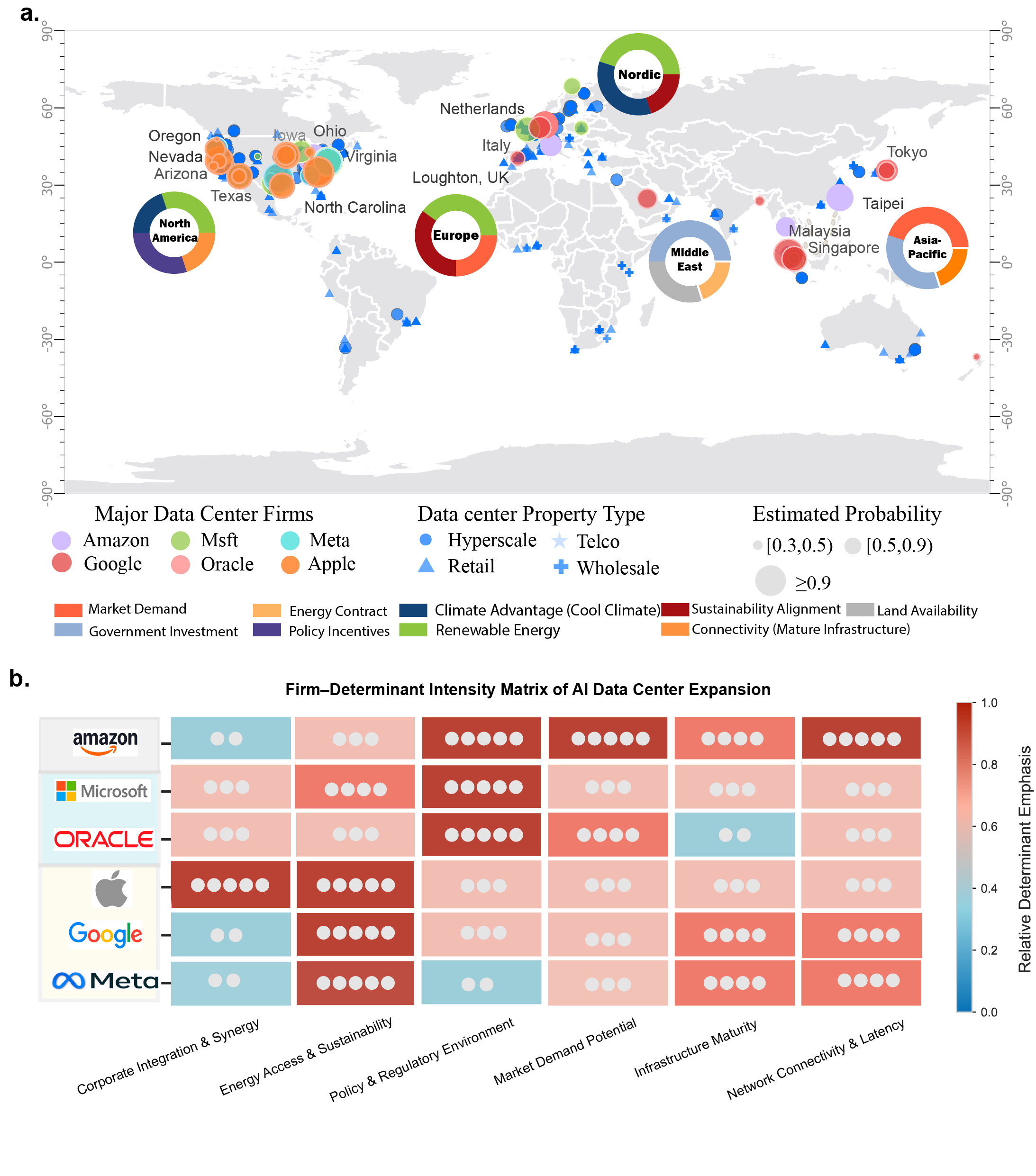}
\caption{(a) Global distribution of data-center sites showing the transition from legacy to AI-specific infrastructure. Small blue markers represent pre-existing enterprise and cloud data centers, while larger color-coded circles denote newly identified AI data centers with varying estimated probabilities of siting. The ringed overlays summarize dominant regional determinants.(b) Firm–determinant intensity heatmap of AI data-center expansion. Color gradients represent the relative importance of the estimated key siting factors. }\label{fig1}
\end{figure}

Figure~\ref{fig1} illustrates the global spatial evolution and underlying determinants of AI data-center siting. 
The smaller blue markers represent legacy cloud and enterprise facilities, whereas the larger color-coded circles denote newly identified AI-specific data centers derived from large language model (LLM) classification. 
The contrast between these distributions reveals a structural transition from historically dispersed siting—previously guided by user proximity and latency—to the emergence of geographically concentrated clusters of AI infrastructure. 
These clusters are most pronounced in North America, Europe, and the Asia–Pacific, which together host over 90\% of global projected AI compute capacity among leading firms. 
The dominant corridors align with regions characterized by abundant renewable-energy potential, mature grid and fibre-optic networks, favorable climatic conditions for cooling, and stable policy frameworks\citep{mavani2024artificial,Zhiwei2022,HYVONEN2024114619}. 
This convergence underscores how energy availability, institutional governance, and digital-connectivity advantages jointly shape the geography of next-generation computational infrastructure.

Distinct regional configurations highlight the heterogeneity of siting determinants. 
North America functions as a renewable-energy and policy corridor, where extensive solar and wind resources, efficient cooling conditions, and long-standing fiscal incentives sustain dense hyperscale clusters in Virginia, Texas, Ohio, and North Carolina. 
Europe forms a regulated green-energy hub centered on the Netherlands, the United Kingdom, and Italy, combining ESG-aligned policy frameworks with high enterprise-cloud demand, while the Nordic region anchors a low-carbon AI-computing zone supported by hydro and wind resources and naturally cool climates \citep{HYVONEN2024114619}. 
In contrast, the Asia–Pacific region exhibits a policy-driven and demand-led pattern, reflecting rapid AI adoption, state-supported infrastructure investment, and strong government coordination in Singapore, Taipei, Malaysia, and Japan. 
The Middle East represents an emerging sovereign AI frontier characterized by fiscal incentives, low land-competition environments, and state-led digital-economy diversification.

Firm-level differentiation further reinforces these spatial dynamics (Figure~\ref{fig1}b). 
Amazon exhibits the broadest geographic diversification, scaling with market demand and infrastructure maturity across regions. 
Microsoft and Oracle expand primarily along regulatory and tax-incentive corridors, while Google and Meta anchor their siting strategies around renewable-energy corridors and high-connectivity nodes to minimize latency and grid intensity. 
Apple, by contrast, adopts a vertically integrated strategy, concentrating within renewable-powered U.S. grids to optimize operational efficiency and alignment with its manufacturing and service ecosystems. 
These configurations can be synthesized into five strategic archetypes: renewable-anchored, policy-responsive, infrastructure-consolidating, market-expanding, and vertically integrated AI ecosystems—each reflecting distinct trade-offs between energy availability, policy alignment, and network reach.

Together, these patterns indicate a bifurcated trajectory in the global evolution of AI infrastructure. 
Hyperscale cloud providers are pursuing an expansion-oriented pathway emphasizing market coverage, latency reduction, and rapid scaling, while vertically integrated ecosystem firms follow an efficiency-oriented pathway focused on renewable coupling and operational precision. 
This divergence suggests that the geography of AI compute is becoming a structural element of regional electricity systems—concentrating load growth, reshaping grid utilization, and altering the spatial distribution of energy demand. 
As the clustering of AI facilities deepens, the intersection of digital and energy infrastructures will increasingly determine the pace and sustainability of the global energy transition.

\subsection{Electricity consumption projection: rapid scaling of AI data centers elevates system-level energy demand}
\begin{figure}[H]
\centering
\includegraphics[width=0.9\textwidth]{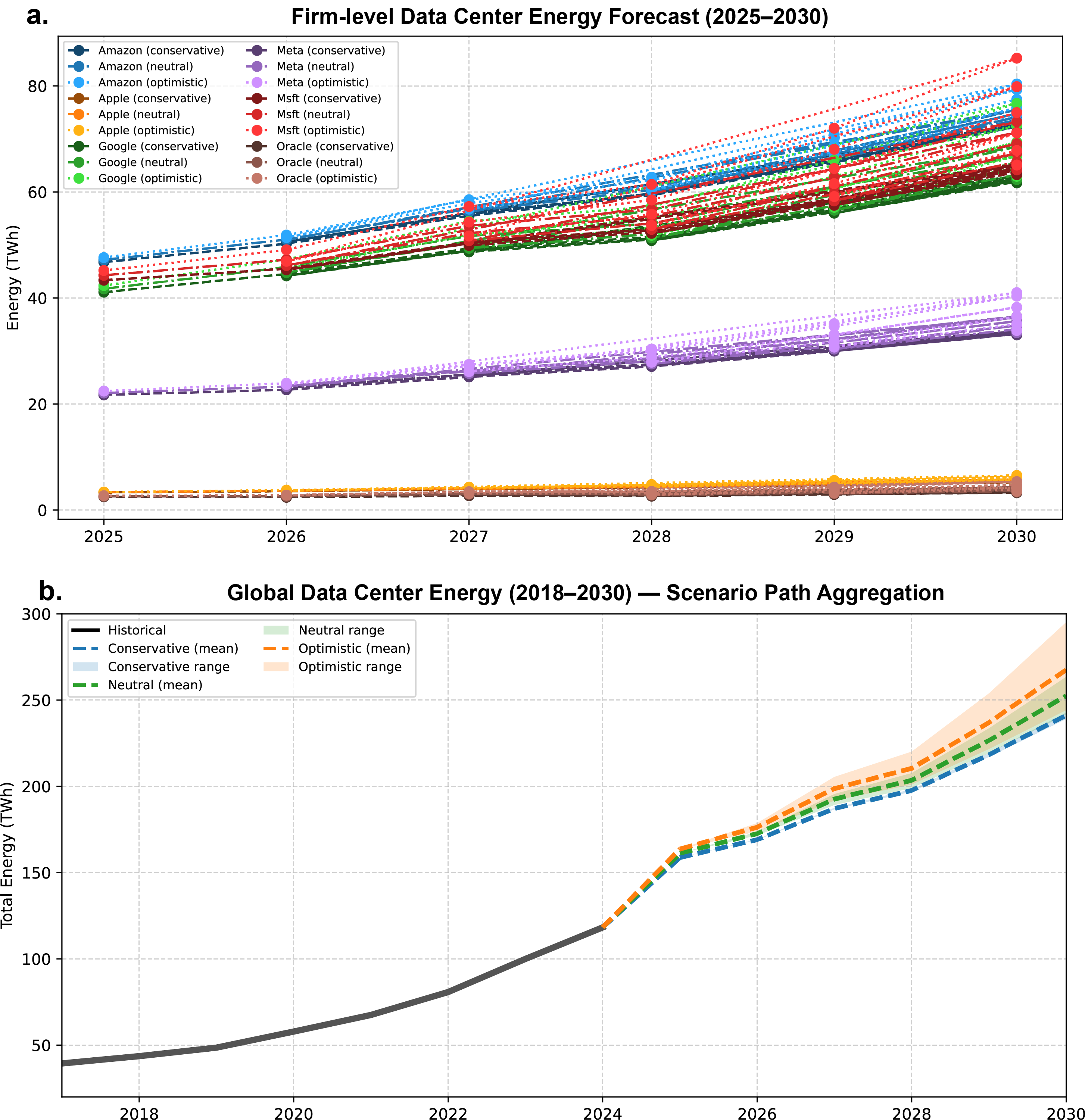}
\caption{(a) Firm-level electricity consumption of data centers from 2025 to 2030 under three site-expansion scenarios. Each trajectory is derived from firm-specific expansion rates (15\%, 25\%, and 35\% for conservative, neutral, and optimistic scenario, respectively) combined with the projected share of AI-intensive workloads by firm. 
(b) Aggregated global electricity consumption of data centers from 2018 to 2030, showing historical trends and scenario-based forecasts. The solid grey line represents historical values, while the colored dashed lines denote the mean and range across the three scenarios. }\label{fig2}
\end{figure}

Figure 2 presents the projected electricity consumption of AI data centers under three growth scenarios from 2025 to 2030. 
Figure 2a illustrates firm-level trajectories for Amazon, Microsoft, Google, Meta, Oracle, and Apple across conservative, neutral, and optimistic assumptions. 
The energy demand forecasts for all firms show consistent upward trends, reflecting the intensification of AI workloads and the proliferation of large-model training clusters. 
Aggregate electricity use among hyperscale operators is projected to rise by approximately 40–70\% over the period, though the pace and scale vary considerably by firm. 
Amazon, Microsoft, and Google are forecasted to maintain the highest absolute demand, consistent with their extensive cloud-service portfolios and broad geographic reach. 
Meta follows comparable but more moderate forecast paths, supported by continuous improvements in model efficiency and renewable procurement strategies \citep{Synergy2024,Uptime2024,AzureAI2024,GoogleCFE2023,Meta2024_10K,AWS_ReInvent2023}. 
Apple and Oracle exhibit lower projected energy demand, aligning with their smaller operational footprints and vertically integrated architectures. 
Across all firms, the optimistic scenario accelerates notably after 2027, coinciding with large-scale deployment of generative-AI infrastructure and training workloads\citep{Apple2024_10K,OCI_NVIDIA2024}.

Figure 2b aggregates these firm-level forecasts to depict the global evolution 
of electricity demand associated with leading AI operators from 2018 to 2030. 
The solid gray line represents historical estimates derived from industrial disclosures 
and utility data, while the colored dashed lines correspond to scenario-based projections. 
Total electricity consumption by these leading firms’ data centers is expected to rise 
from approximately 118~TWh in 2024 to between 239~TWh (conservative) and 295~TWh (optimistic) 
by 2030, implying a compound annual growth rate (CAGR) of roughly 
13-17\%. This increase is equivalent to adding the electricity demand 
of a medium-sized national market every two years, or comparable to the annual electricity 
use of about 27~million U.S. households. The projected magnitude is quantitatively consistent 
with the International Energy Agency’s (IEA) forecast that total global data center electricity 
demand will reach around 945 TWh by~2030 (approximately 3\% of global power consumption) \citep{IEA2025_EnergyAI}, compared with 
about 1.5\% in~2024. These projections collectively illustrate the accelerating energy 
footprint of AI infrastructure within the broader global power system.

Taken as a whole, these projections demonstrate that AI-driven digital infrastructure is transitioning from a marginal load to a structurally significant component of the global power system. 
Even under conservative assumptions, AI data centers are emerging as major contributors to electricity demand growth within national and regional grids. 
The accelerated expansion observed in the optimistic scenario underscores the urgency of integrating digital-load planning into power-system forecasting, renewable siting, and investment strategies. 
Absent substantial improvements in model efficiency, cooling systems, and power-usage effectiveness, the cumulative demand from AI computation could offset a sizeable portion of planned renewable additions in advanced economies by the end of the decade. 
This inflection marks the point at which the evolution of digital infrastructure becomes an active determinant of electricity-system design, reinforcing the need for coordinated strategies linking computational growth, renewable deployment, and carbon-mitigation policy.

\subsection{Regional power-stress implications of AI data-center clustering}
\begin{figure}[H]
\centering
\includegraphics[width=0.9\textwidth]{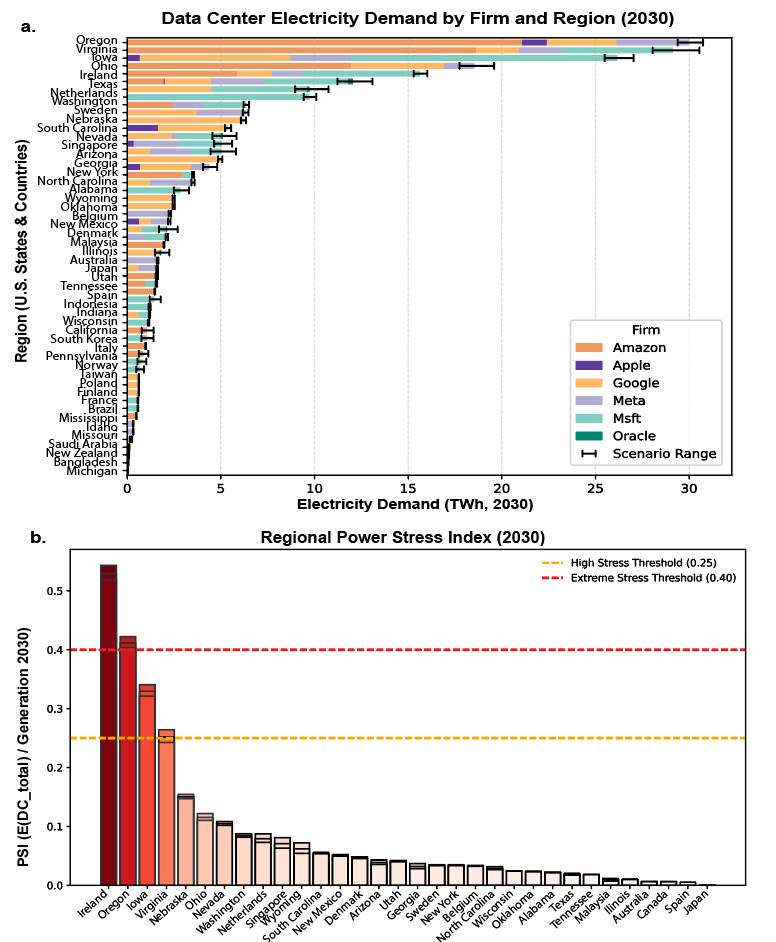}
\caption{(a) Regional distribution of AI data-center electricity demand by firm in 2030, combining U.S. states and overseas host countries. Error bars denote scenario uncertainty across conservative, neutral, and optimistic scenarios. (b) Regional Power Stress Index (PSI) in 2030, defined as the ratio of data center electricity demand to total regional generation capacity. Dashed horizontal lines indicate stress thresholds of 0.10, 0.25, and 0.40 corresponding to industry reliability benchmarks reported by the International Energy Agency, EPRI, and IEEE. Values above 0.25 represent regions where industrial digital loads may require dedicated transmission expansion or peak-support resources, while values above 0.40 indicate extreme concentration of demand relative to local generation.}\label{fig3}
\end{figure}

Figure~\ref{fig3} evaluates the regional electricity implications of AI data-center expansion by 2030. 
Figure 3a displays the projected electricity demand by firm and region, revealing a highly uneven spatial distribution of AI-related loads. 
Fewer than ten regions account for nearly two-thirds of total projected demand, illustrating a large concentration of infrastructure in a limited number of states and countries. 
Oregon, Virginia, Iowa, Texas, and Ireland emerge as dominant hubs, each exceeding 15~TWh of annual demand under the neutral scenario. 
These regions combine favorable attributes—mature hyperscale clusters, abundant renewable resources, and established regulatory or policy incentives—that reinforce their attractiveness for large-scale AI operations. 
Second-tier regions such as the Netherlands, Washington, and Singapore display moderate but rapidly increasing loads, often driven by single-firm expansions. 
Scenario ranges indicate that uncertainty in deployment pace and firm strategy could alter local electricity demand by up to 30\%, highlighting the challenge of forecasting and accommodating digital infrastructure in regional power planning.

Figure 3b quantifies these spatial imbalances using the Power Stress Index (PSI), defined as the ratio of data-center electricity demand to total projected generation capacity in 2030\citep{IEA2021_ElectricitySecurity}. 
Regions such as Ireland and Oregon already surpass the high-stress threshold of 0.25, with Ireland approaching 0.5, implying that AI-related electricity consumption could absorb nearly half of local generation. 
Virginia, Nebraska, and Washington also exhibit elevated PSI levels between 0.20 and 0.30, reflecting the concentration of multiple hyperscale campuses within transmission-constrained zones. 
In contrast, regions with larger and more diversified power systems—including Texas, Japan, and Australia—maintain PSI values below 0.10, indicating greater absorptive capacity and structural resilience. For instance, Texas exhibits structural advantage, with independent grid governance, substantial renewable potential, competitive electricity prices, and efficient interconnection framework, which enable the integration of incremental AI-related loads while supporting continued renewable expansion. 
As national electricity demand is projected to grow at approximately 2.4\% per year through 2030—driven largely by AI infrastructure—such regions are well positioned to capture the economic benefits of digital expansion at comparatively lower system cost. 
This cross-regional heterogeneity underscores that the geography of AI data-center clustering not only amplifies spatial supply–demand asymmetries but also delineates a new divide between regions exposed to power stress and those structurally advantaged to leverage digital–energy integration for sustainable growth.

Taken together, these results highlight an emerging spatial asymmetry between regions experiencing power-system stress and those structurally capable of accommodating AI-driven demand. 
High-stress regions such as Oregon, Iowa, and Ireland face renewable intermittency, transmission constraints, and rising marginal emissions, whereas resilient systems like Texas illustrate how institutional flexibility and diversified resources mitigate such pressures. 
Absent coordinated planning among utilities, regulators, and data-center developers, these disparities may intensify price volatility and hinder decarbonization efforts\citep{DOERING2021}. 
Conversely, integrating AI infrastructure siting into national energy strategies could enable renewable-rich, low-cost regions to serve as anchors for grid decarbonization and system flexibility. 
This evolving spatial divergence thus embodies both a systemic risk and a strategic opportunity for aligning digital expansion with sustainable electricity transitions\citep{BOLONCANEDO2024}.

\section{Discussion}
The integrated results demonstrate that AI-driven data-center expansion is emerging as a structurally significant force within the global electricity system \citep{LAMBERT2026107968,Stojkovic2025}.
Across firms and regions, the analysis reveals a coherent sequence: geographically concentrated siting patterns, rapidly accelerating electricity demand, and emerging regional power stress. 
These findings indicate that the digital infrastructure underpinning AI innovation is no longer exogenous to energy-system dynamics. 
Instead, it has become an endogenous determinant of grid configuration, investment priorities, and decarbonization pathways\citep{BOLONCANEDO2024}.

The clustering of AI data centers in a limited number of favorable regions—primarily in North America, Western Europe, and parts of the Asia–Pacific—creates pronounced spatial asymmetries in load distribution. 
High-stress regions such as Oregon, Virginia, and Ireland face the convergence of rapid AI expansion, renewable intermittency, and transmission bottlenecks, generating new points of vulnerability in balancing capacity. 
Conversely, systems with more diversified resources and institutional flexibility, such as Texas, are better positioned to integrate AI-driven loads efficiently and at lower cost. 
The large projected electricity demand thus represents both a systemic challenge and a strategic opportunity. 
Without coordinated planning, concentrated AI demand could delay renewable integration, reinforce fossil capacity, and intensify price volatility. 
However, if strategically integrated, data centers could serve as anchor loads for renewable Power Purchase Agreements (PPAs), flexible off-takers for variable generation, and hubs for co-located storage and demand response\citep{de2025ai,davenport2024ai}. 
This divergence highlights how governance, market design, and spatial resource endowment jointly shape the distributional consequences of digital electrification.

At the policy level, these findings underscore the need for anticipatory frameworks that explicitly incorporate AI-related demand into national energy modelling and capacity-expansion planning. 
Conventional forecasts that treat digital infrastructure as exogenous commercial load risk underestimating peak capacity requirements and the spatial concentration of new demand. 
Integrating digital-load projections into long-term planning would enable regulators and utilities to align renewable siting, storage investment, and transmission reinforcement with emerging AI growth corridors. 
Fiscal and zoning incentives to attract digital investment should be complemented by requirements for renewable procurement, locational diversity, and grid-contribution mechanisms. 
In emerging markets, policy design must balance the benefits of digital infrastructure development with the risk of exacerbating power scarcity and regional inequality. 
Subnational regulators may also need to strengthen interregional grid interties and distributed-storage policies to mitigate electricity price volatility and localized carbon intensity.

More broadly, the co-evolution of digitalization and electrification marks a conceptual turning point in energy-system governance\citep{MLSYS2022_462211f6}.  AI infrastructure is transforming electricity demand from a passive outcome of economic activity into an active determinant of system configuration. 
Its social and economic value increasingly depends on the sustainability of its power source and its integration within low-carbon grids.
Future research should quantify the feedback loops between AI-compute trajectories, carbon intensity, and investment flows across the energy–digital interface. 
Understanding and governing this coupling between computational growth and electricity systems will be central to achieving a resilient, equitable, and low-carbon digital economy.
\bibliography{sn-bibliography}

\section*{Methods}\label{sec11}
This study develops a hybrid Retrieval-Augmented Generation (RAG) + Large Language Model (LLM) forecasting framework to predict the geographic and operational evolution of hyperscale data centers operated by major technology companies. 
The framework integrates structured document retrieval, language sentiment analysis, and LLM-based contextual reasoning to infer (i) potential expansion locations, (ii) baseline technical parameters, and (iii) multi-year energy trajectories.

The workflow consists of five main stages: 
(1) construction of a RAG knowledge base from multi-year corporate and regional sources, 
(2) sentiment-aware identification of likely future data center locations, 
(3) retrieval of location-specific contextual information, 
(4) extraction of baseline operational parameters for the current year, 
(5) multi-year forecasting of key technical and environmental metrics using an LLM-guided model and physically interpretable energy equations, and 
(6) construction of Power Stress Index (PSI) that compares projected data center demand against the available regional electricity supply capacity.
\begin{figure}[H]
\centering
\includegraphics[width=0.8\textwidth]{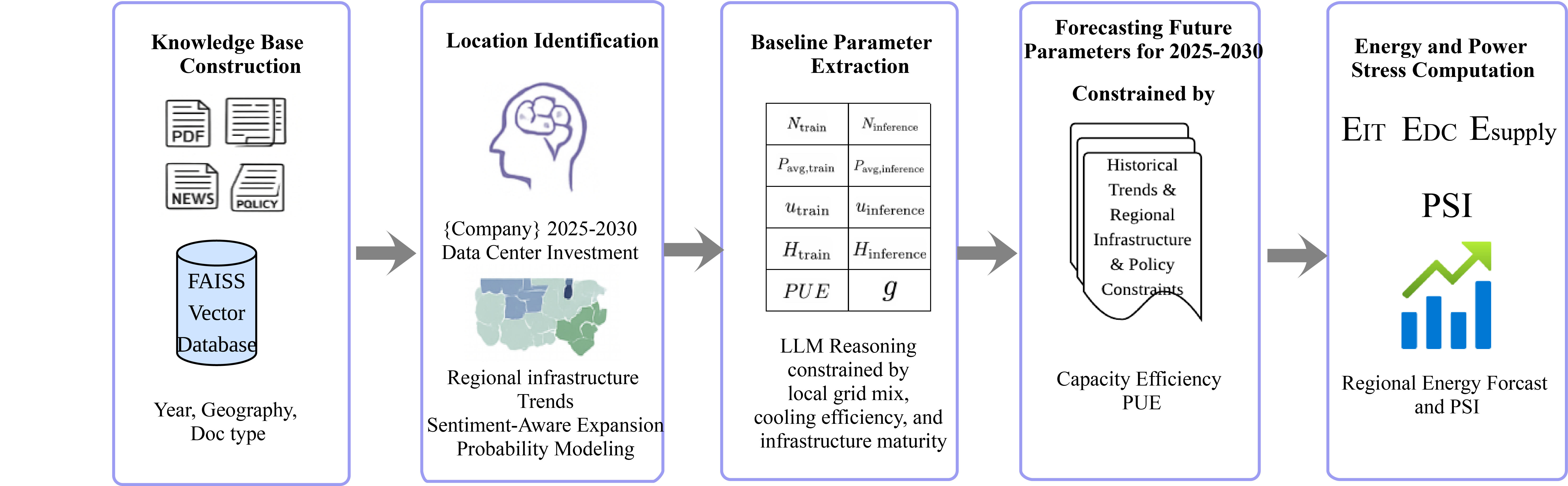}
\caption{The Workflow Diagram}\label{Diagram}
\end{figure}
\subsection*{1) Knowledge Base Construction}
A multi-source text corpus was constructed for each firm, including press releases, annual earnings reports, environmental sustainability disclosures, infrastructure investment announcements, government policy documents, and major news articles from 2015--2025. 
All texts were preprocessed and embedded using the Hugging Face \texttt{all-MiniLM-L6-v2} model, then indexed in a \texttt{FAISS} (Facebook AI Similarity Search) vector database for semantic retrieval. 
Each entry retained metadata such as publication date, geographic reference, and document type, enabling targeted retrieval (e.g., ``AWS Taiwan 2025 expansion energy infrastructure'').

\subsection*{2) Location Identification via RAG + LLM}

To infer likely future expansion sites between 2025 and 2030, the framework explicitly prompts the LLM to \textit{predict which geographic regions a firm is most likely to expand into}.  
This prediction is based on contextual evidence retrieved from the RAG knowledge base, which includes firm sustainability reports, capital expenditure statements, and government or media coverage.  

The RAG system is first queried using firm-specific and temporal prompts, such as \texttt{\{firm\} 2025--2030 data center investment or construction plans}.  
The retrieved documents contain both qualitative statements (e.g., ``strong regional demand growth,'' ``strategic infrastructure partnership'') and quantitative indicators (e.g., ``\$2B allocated to APAC expansion'').  

Each document is analyzed by the LLM not only for content relevance but also for its \textbf{sentiment and linguistic tone}.  
Positive sentiment toward investment, expansion, or infrastructure development—particularly in earnings calls or annual reports—acts as an implicit prior that increases the model’s belief in a region’s likelihood of receiving new data center capacity.  
Conversely, negative sentiment (e.g., mentions of cost control, divestment, or policy uncertainty) decreases the associated probability weight.  

The LLM then synthesizes these cues into a structured probability distribution over candidate locations, producing outputs of the form:
\[
P(\text{expansion at site } i \mid \text{evidence}) = f_{\text{LLM}}(\text{retrieved text embeddings}, \text{sentiment scores})
\]
This probabilistic representation is used in later stages to weight forecasts of capacity growth and energy consumption at each site.

\subsection*{3) Baseline Parameter Extraction}
For each confirmed or high-probability site, the system used a RAG-conditioned LLM prompt to extract \textbf{current-year baseline parameters} describing the data center’s IT and facility operations. 
The model was instructed to reason explicitly about local grid mix, cooling efficiency, and infrastructure maturity, and to avoid copying assumptions between regions. 
The extracted parameters included:
\[
\{ N_{\text{train}}, N_{\text{inference}}, P_{\text{avg,train}}, P_{\text{avg,inference}}, u_{\text{train}}, u_{\text{inference}}, H_{\text{train}}, H_{\text{inference}}, PUE, g \}
\]
All outputs were formatted as structured JSON (JavaScript Object Notation) objects, accompanied by short notes summarizing assumptions and references. Each variable corresponds to a measurable operational attribute of the local compute fleet.

\begin{threeparttable}
\small
\caption{Definition of baseline parameters used in the GPU-hours-based formulation.}
\begin{tabularx}{\textwidth}{l l l X}
\toprule
\textbf{Symbol} & \textbf{Definition} & \textbf{Units} & \textbf{Description} \\
\midrule
$N_{\text{train}}$ & Number of accelerators for AI training & count & GPUs/TPUs/ASICs simultaneously engaged in training workloads. \\
$N_{\text{inference}}$ & Number of accelerators for inference & count & Active inference accelerators allocated in the site cluster. \\
$P_{\text{avg,train}}$ & Average power draw per training accelerator & kW & Mean electrical power while training, including memory and I/O overheads. \\
$P_{\text{avg,inference}}$ & Average power draw per inference accelerator & kW & Mean electrical power during inference mode. \\
$u_{\text{train}}$ & Average utilization factor (training) & – & Fraction (0–1) of wall-clock time each accelerator performs active computation. \\
$u_{\text{inference}}$ & Average utilization factor (inference) & – & Typically lower (0.2–0.4) due to bursty traffic and idle periods. \\
$H_{\text{train}}$ & Aggregate training hours per accelerator in period $t$ & h & Derived from duty cycle multiplied by period length. \\
$H_{\text{inference}}$ & Aggregate inference hours per accelerator in period $t$ & h & Operating hours of inference deployment in the same period. \\
$PUE$ & Power Usage Effectiveness & – & Ratio of total facility energy to IT load; adjusted by operator and latitude model. \\
\bottomrule
\end{tabularx}
\end{threeparttable}
\newpage

\subsection*{4) Forecasting Future Parameters}
A second RAG + LLM chain projected the time evolution of these parameters over a five-year horizon. 
The prompt incorporated both historical operational trends and retrieved regional information, constraining LLM reasoning with empirically observed hyperscale patterns: 10--20\% annual capacity growth, 1--3\% annual efficiency gains, and gradual improvements in utilization.
The model generated structured forecasts for each year $t \in [2025,2030]$, each annotated with explanatory notes and relevant document sources.

\subsection*{5) Energy Computation}
Forecasted operational parameters were post-processed in Python to derive the physical energy demand of each site. 
The \textbf{IT load energy} ($E_{\text{IT}}$) represents the direct electrical consumption of computing equipment, while the estimated AI \textbf{data center energy demand} ($E_{\text{DC}}$) includes both IT and non-IT overheads such as cooling and power distribution losses.

\begin{align}
E_{\text{IT}}(t) &= \frac{N_{\text{train}} P_{\text{avg,train}} u_{\text{train}} H_{\text{train}} + N_{\text{inference}} P_{\text{avg,inference}} u_{\text{inference}} H_{\text{inference}}}{1000}, \\
E_{\text{DC}}(t) &= PUE(t) \times E_{\text{IT}}(t)
\end{align}

Here, $N$, $P_{\text{avg}}$, $u$, and $H$ denote the number of accelerators, average power draw (kW), utilization factor, and annual operating hours, respectively.  
$E_{\text{IT}}$ is thus measured in MWh, and $E_{\text{DC}}$ scales this load by the facility’s Power Usage Effectiveness ($PUE$), capturing site-specific cooling and infrastructure efficiency.  

\subsection*{6) Implementation}
The full pipeline was implemented in Python using the \texttt{LangChain} framework for composable RAG pipelines, \texttt{FAISS} for vector retrieval, and the \texttt{OpenAI GPT-4o-mini} model as the reasoning component (temperature = 0.3). 
All modules were orchestrated via reproducible runnables with strict JSON parsing for downstream numerical analysis.

\subsection*{7) Sentiment-Aware Expansion Probability Modeling}
The sentiment analysis stage produced, for each candidate region $i$, an \textbf{expansion likelihood score} $P_i$ defined as:
\begin{equation}
P_i = \frac{S_i \cdot R_i}{\sum_j S_j \cdot R_j}
\label{eq:P_i}
\end{equation}
where 
$S_i$ is the normalized sentiment score for region $i$ (ranging from $-1$ to $+1$), and 
$R_i$ is the retrieval relevance score (cosine similarity in the embedding space). 
These probabilities informed the sampling or prioritization of forecast targets in subsequent steps. 
Regions frequently mentioned in corporate reports \emph{and} described with positive financial tone thus received elevated probabilities of selection.

This coupling of semantic retrieval and sentiment tone weighting enables the model to capture \textbf{implicit strategic intent} in corporate communications—an important predictor of near-term data center expansion behavior.

By fusing semantic retrieval, sentiment weighting, and physical energy modeling, this hybrid framework bridges qualitative corporate signals and quantitative operational forecasting. 
The approach captures how firms’ language choices in financial and environmental disclosures correlate with real-world infrastructure trajectories, enabling a more interpretable and evidence-grounded forecast of global data center development.

\subsection*{8) Scenario design and aggregation}

To capture uncertainty in both firm-level expansion intensity and technological efficiency, we constructed three forward scenarios $s$-\textit{conservative}, \textit{neutral}, and \textit{optimistic}-for 2025–2030. Each scenario jointly controls (i) the rate of new AI data-center additions and (ii) the efficiency gains of all operating sites. The annual growth rate of new AI load is set to $g_{\text{new},s} \in \{0.15,\,0.25,\,0.35\}$ for scenario s among the three scenarios, while existing stock consumption for each firm grows at $g_{\text{stock}} = 0.10$. Firm-specific AI shares $p_{\mathrm{AI}}(f,t)$ increase gradually over time to reflect the rising share of AI workloads in total compute demand.
Firm-specific AI shares $p_{\mathrm{AI}}(f,t)$ were defined to represent the evolving proportion of AI-related compute workloads within each operator’s total data-center activity. 
For each firm $f$, $p_{\mathrm{AI}}(f,t)$ increases over time according to a discrete schedule calibrated to reflect the firm’s relative intensity of AI adoption between 2025 and 2030:
\[
p_{\mathrm{AI}}(f,t) \in \{p_{1}(f), p_{2}(f), p_{3}(f)\},
\]
where $p_{1}(f)$, $p_{2}(f)$, and $p_{3}(f)$ correspond to the firm’s baseline (2025), mid-phase (2026–2027), and mature-phase (2028–2030) AI workload shares, respectively. 
Formally, this can be expressed as:
\[
p_{\mathrm{AI}}(f,t) =
\begin{cases}
p_{1}(f), & t = 2025,\\[4pt]
p_{2}(f), & t \in \{2026,\,2027\},\\[4pt]
p_{3}(f), & t \in \{2028,\,2029,\,2030\},\\[4pt]
p_{3}(f), & t > 2030.
\end{cases}
\]

The parameter set $\{p_{1}(f),p_{2}(f),p_{3}(f)\}$ is firm-specific, capturing heterogeneity in AI adoption trajectories across major hyperscale operators:
\[
\begin{aligned}
&\text{Amazon: } (0.30,\,0.40,\,0.60), \quad
\text{Apple: } (0.25,\,0.30,\,0.35), \\
&\text{Google: } (0.35,\,0.40,\,0.60), \quad
\text{Meta: } (0.35,\,0.50,\,0.60), \\
&\text{Microsoft: } (0.35,\,0.45,\,0.60), \quad
\text{Oracle: } (0.25,\,0.35,\,0.50).
\end{aligned}
\]

This stepwise formulation approximates each firm’s increasing allocation of computational resources to AI model training and inference. 
The heterogeneity in $p_{\mathrm{AI}}(f,t)$ reflects differences in investment magnitude, infrastructure build-out, and strategic focus among major operators. 
Firms with large-scale and early commitments to generative AI—such as Amazon, Google, Meta, and Microsoft—exhibit steeper trajectories, consistent with multibillion-dollar data-center expansions, dedicated AI accelerator deployment (e.g., Trainium, H100, TPUv5), and vertically integrated AI ecosystems. 
By contrast, vertically integrated or enterprise-focused operators such as Apple and Oracle show slower growth, consistent with smaller-scale AI infrastructure investment, reliance on consumer-device optimization rather than large-model training, and a later pivot toward cloud-based AI services. 
Accordingly, the evolution of $p_{\mathrm{AI}}(f,t)$ captures each firm’s position along the spectrum of AI-driven digital infrastructure transformation, serving as a proxy for the relative share of compute capacity devoted to AI workloads within their overall fleet.

For each firm $f$, the existing stock electricity consumption evolves as
\begin{equation}
E^{\text{stock}}_{f,t} = E^{\text{stock}}_{f,2024}\,(1+g_{\text{stock}})^{t-2024}.
\label{eq:estock}
\end{equation}

New-site electricity demand follows compounding growth of AI workloads and the firm-specific AI share schedule:
\begin{align}
E^{\text{AI,new}}_{f,t,s} &= E^{\text{AI,new}}_{f,t-1}\,[1+g_{\text{new}}(s)], \label{eq:eain}\\
E^{\text{new}}_{f,t,s} &= \frac{E^{\text{AI,new}}_{f,t,s}}{p_{\mathrm{AI}}(f,t)}. \label{eq:enew}
\end{align}

The total firm-level electricity demand is then
\begin{equation}
E^{\text{tot}}_{f,t,s} = E^{\text{stock}}_{f,t,s} + E^{\text{new}}_{f,t,s}.
\label{eq:etot}
\end{equation}

To represent uncertainty arising from multiple independent siting pathways, we simulate a set of $P_{t,s}$ plausible firm–region configurations for each year $t$ and scenario $s$.  
Global electricity demand in year $t$ and scenario $s$ is the ensemble mean:
\begin{equation}
\overline{E}_{t,s} = \frac{1}{P_{t,s}} \sum_{p=1}^{P_{t,s}} E^{\text{tot}}_{f,t,s,p},
\label{eq:emean}
\end{equation}
and the range 
\[
[\min_p E^{\text{glob}}_{t,s,p}, \max_p E^{\text{glob}}_{t,s,p}]
\]
defines the shaded uncertainty bands in the scenario plots, 
where $E^{\text{glob}}_{t,s,p}$ denotes the global total electricity demand 
in year~$t$ under scenario~$s$ for projection path~$p \in \{1,\ldots,P_{t,s}\}$. 
The 2024 observed total serves as the common anchor point for all trajectories.

\subsection*{9) Regional allocation of energy}

Firm-level electricity demand is spatially distributed according to two complementary weights: (i) AI siting probabilities derived from the LLM siting model, and (ii) historical stock weights from the existing data-center inventory.

For AI-related new sites, firm–year location weights are proportional to modelled AI energy at each location $\ell$:
\begin{align}
    w^{\text{AI}}_{f,t,\ell} = \frac{E^{\text{AI,loc}}_{f,t,\ell}}{\sum_{\ell'} E^{\text{AI,loc}}_{f,t,\ell'}}.
\end{align}
Here $E^{AI,\text{loc}}_{f,t,\ell}$ denotes the modelled AI-related electricity 
demand assigned to firm~$f$ at location~$\ell$ in year~$t$, as determined from 
the LLM-based siting model and corresponding regional AI workload intensity.

For legacy and non-AI sites, we use historical weights based on observed facility counts. For legacy and non-AI sites, we use historical weights based on observed facility counts\footnotemark[1]:
\begin{align}
    w^{\text{hist}}_{f,\ell} = 
\frac{\mathrm{siteCount}_{f,\ell}}{\sum_{\ell'} \mathrm{siteCount}_{f,\ell'}}.
\end{align}

Let $r(\ell)$ map locations to regions. Regional electricity use for firm~$f$ and scenario~$s$ is:
\begin{align}
E_{f,t,s,r} &=
\sum_{\ell:\,r(\ell)=r} \Big[
w^{\text{AI}}_{f,t,\ell}\,E^{\text{AI,new}}_{f,t,s} +
w^{\text{hist}}_{f,\ell}\big(E^{\text{nonAI,new}}_{f,t,s} + E^{\text{stock}}_{f,t}\big)
\Big].
\end{align}
\footnotetext[1]{\textit{Legacy (historical) data-center counts} refer to the number of distinct operating campuses under each firm’s control as of end-2024, including both owned and long-term leased facilities. These are derived from the observed 2015–2024 dataset and serve as proxies for baseline spatial capacity prior to the AI expansion. Since 2024, the global data-center landscape has entered a rapid expansion phase driven by generative-AI workloads, with AI infrastructure investment reaching a record \$57 billion globally\footnote{DataCenter Dynamics (2025), \emph{AI drove record \$57 bn in data-center investment in 2024}.}. This ``AI data-center boom'' marks a structural inflection in digital-infrastructure growth.}
Summing over all firms yields regional totals:
\begin{align}
E^{\text{region}}_{t,s,r} = \sum_f E_{f,t,s,r}.
\end{align}
This approach preserves firm-level energy consistency while ensuring that AI-driven growth concentrates in high-probability siting regions identified by the LLM model, and that legacy loads follow empirically observed footprints.
\subsection*{10) Cross-validation}
Aggregate electricity consumption by the six leading hyperscale operators is projected to rise from approximately 118~TWh in 2024 to 239–295~TWh by 2030. This projection is broadly consistent with the International Energy Agency’s (IEA) estimate that global data-center electricity demand will reach roughly 945~TWh by 2030 (IEA, 2025). To benchmark the modelled values, we derive an implied 2030 consumption level for the six leading firms based on the IEA’s global forecast. Assuming (i) that hyperscale data centres account for 70\% of global data-centre electricity use (Synergy Research Group, 2024), and (ii) that the top six operators collectively represent 40\% of hyperscale activity, the corresponding implied electricity consumption is:

\[
E^{\text{six-firms, IEA-based}}_{2030}
= 945~\text{TWh} \times 0.70 \times 0.40
= 264.6~\text{TWh} \approx 265~\text{TWh}.
\]

This cross-validation demonstrates that the projected range of 239--295~TWh lies well within the values implied by independent international forecasts. The close correspondence reinforces confidence in the representativeness of the modelled global totals and supports the credibility of the upper-bound scenario in light of established external benchmarks. Overall, the modelled range remains broadly aligned with independent global projections and reflects plausible hyperscale market dynamics.


\subsection*{11) Regional Power Stress Index (PSI)}

To assess the potential strain imposed by data center expansion on regional electricity systems, we constructed a \textbf{Power Stress Index (PSI)} that compares projected hyperscale data center demand against the available regional electricity supply capacity.

For each region $r$ and year $t$, the PSI is defined as:
\begin{align}
\text{PSI}_{r,t} = \frac{E^{\text{DC}}_{r,t}}{E^{\text{Supply}}_{r,t}}
\end{align}
where $E^{\text{DC}}_{r,t}$ denotes total data center electricity demand (sum of AI and non-AI workloads), and $E^{\text{Supply}}_{r,t}$ denotes total net electricity generation or capacity available in that region.

\textbf{Regional supply baselines} ($E^{\text{Supply}}_{r,2024}$) were obtained from U.S. Energy Information Administration (EIA) and IEA datasets covering 2020–2024, expressed in terawatt-hours (TWh). 
Electricity generation data were extracted from \texttt{global\_his.xlsx}, interpolated where necessary, and extrapolated to 2030 using a compound annual growth rate (CAGR):
\begin{align}
E^{\text{Supply}}_{r,t} = E^{\text{Supply}}_{r,2024} \times (1 + \text{CAGR}_r)^{(t-2024)}
\end{align}
with $\text{CAGR}_r = \left(\frac{E^{\text{Supply}}_{r,2024}}{E^{\text{Supply}}_{r,2019}}\right)^{1/5} - 1$.

For U.S. states, generation data were obtained from the U.S. Energy Information Administration (EIA) state electricity profiles. 
All capacities were normalized to ensure consistency between regional boundaries used in the RAG siting model and the statistical reporting boundaries in the energy datasets.

Regions with $\text{PSI}_{r,t} > 0.25$ were identified as potentially exposed to moderate stress, while $\text{PSI}_{r,t} > 0.40$ indicates a high-stress condition where concentrated hyperscale loads could require dedicated transmission or peak-support investments. 
These thresholds align with reliability guidelines reported by the International Energy Agency (IEA) and the Electric Power Research Institute (EPRI), which note that industrial loads exceeding 25–30\% of generation capacity may materially affect system reliability and reserve margins.

Spatially, PSI values were visualized as heat maps and binned into quantiles to highlight emerging AI infrastructure corridors—regions where concentrated expansion coincides with constrained supply growth. 
The PSI framework thus translates projected data center energy trajectories into interpretable indicators of grid stress and regional resilience, bridging firm-level forecasts with systemic energy planning concerns.

\section*{Data sources}



The analysis combines multiple proprietary and public datasets to characterize the spatial, technical, and energetic dimensions of AI‐driven data‐center expansion.

\textbf{(1) The LLM corpus} forms the foundational dataset for identifying and characterizing AI data-center sites. It integrates heterogeneous information sources spanning corporate disclosures, energy infrastructure databases, geospatial imagery, and policy documents. Corporate and infrastructure data were obtained from the \textit{SEC EDGAR filings} (\url{https://www.sec.gov/edgar/searchedgar/companysearch.html}), official \textit{ESG and sustainability reports} of major cloud-service providers, and the \textit{Data Center Knowledge Base} (DatacenterDynamics). Energy-related inputs were drawn from the \textit{International Energy Agency (IEA)} and the \textit{U.S. Energy Information Administration (EIA)} datasets, as well as the \textit{Global Power Plant Database} (World Resources Institute) and \textit{OpenStreetMap} for infrastructure topology. Geospatial context was supplemented by \textit{Landsat 8} and \textit{Sentinel-2} satellite imagery retrieved via \textit{Google Earth Engine}, combined with population and land-use layers from the \textit{Global Human Settlement Layer (GHSL)} and \textit{LandScan}. Policy and planning documents were sourced from the \textit{U.S. Department of Energy (DOE)}, the \textit{European Commission Directorate-General for Energy (DG-ENER)}, and national grid operators. The corpus was processed using a fine-tuned large language model to extract structured attributes—location, operational status, capacity, and energy characteristics—and to generate probabilistic siting indicators. These outputs underpin the spatial inference and energy-demand modelling presented in this study.

\textbf{(2) The historical data-center sites} of leading firms—Amazon, Microsoft, Google, Meta, Oracle, and Apple—were obtained from \textit{S\&P Capital IQ} (\url{https://www.capitaliq.com/CIQDotNet/Login-okta.aspx}). Capital IQ provides quarterly updated information on firm facilities, investment timing, and geographic distribution. Each facility record includes attributes such as commissioning year, capacity type, and operational status, allowing for consistent tracking of siting evolution over the past decade. These data were cross-validated with public firm disclosures, environmental filings, and infrastructure registries to ensure temporal and locational accuracy. The resulting dataset enables a longitudinal assessment of firm-specific expansion dynamics and supports the validation of model-derived siting probabilities.

\textbf{(3)The energy capacity data} were assembled from the \textit{International Energy Agency (IEA)} (\url{https://www.iea.org/data-and-statistics}), the \textit{U.S. Energy Information Administration (EIA)} (\url{https://www.eia.gov/electricity/data/state/}), and regional grid operators’ statistical releases. For non-U.S. regions, national electricity and generation capacity data were obtained from the \textit{World Bank Open Data} (\url{https://data.worldbank.org/indicator/EG.ELC.GEN.KH}) and the \textit{OECD Energy Statistics} (\url{https://stats.oecd.org/Index.aspx?DataSetCode=ENERGY_BALANCE}). These datasets provide annual country- and region-level information on net generation, installed capacity, and generation mix. The data were harmonized to 2023 and extrapolated to 2030 using officially reported generation-capacity growth rates and national energy plans. The resulting capacity baselines underpin the computation of the Power Stress Index (PSI) and enable evaluation of regional system resilience under alternative AI-driven demand scenarios.


\newpage
\break
\begin{threeparttable}
\section*{Supplementary Information}

\footnotesize
\setlength{\tabcolsep}{3pt}
\caption{Summary of the six major hyperscale operators, their estimated market shares, and AI-driven data center expansion strategies}
\begin{tabularx}{\textwidth}{L{2.5cm} L{3.2cm} L{2.2cm} L{2.4cm} X}
\toprule
\textbf{firm} & \textbf{Infrastructure scale (2025)} & \textbf{Approx.\ standing} & \textbf{Est.\ market share (2025)} & \textbf{AI expansion strategy} \\
\midrule
Amazon Web Services (AWS) &
32 cloud regions, 102 availability zones (+4 planned) &
Largest operator globally &
30--31\% (global cloud infrastructure) &
Rapid build-out of AI-optimized infrastructure (e.g., Trainium and Inferentia chips); launch of Bedrock and Titan foundation models; dedicated AI clusters for customers. \\

Microsoft Azure &
62 regions, 120 availability zones, 200+ data centers &
Second largest globally &
$\sim$20\% &
Massive AI investment via OpenAI partnership; deployment of NVIDIA H100 GPU clusters and liquid-cooled servers; expansion in Iowa, Virginia, and Sweden. \\

Google Cloud Platform (GCP) &
39 regions, 118 availability zones, 35 owned centers across 10+ countries &
Third largest globally &
$\sim$13\% &
Integration of TPUv5 AI accelerators; target of 24/7 carbon-free energy by 2030; expanding global AI training hubs. \\

Meta Platforms &
24 data-center campuses (\textgreater{}53 million sq ft) &
Leading social-AI infrastructure provider &
$\sim$4--6\% (hyperscale capacity share) &
Refit of compute sites into AI-specific clusters for LLaMA and generative-AI workloads; major CapEx on GPU superclusters; in-house energy-efficient design. \\

Oracle Cloud Infrastructure (OCI) &
46 regions, 56 availability zones &
Fifth largest cloud provider &
$\sim$3--4\% &
Partnership with NVIDIA (2024--2025) for regional AI infrastructure; focus on enterprise-grade AI workloads and secure environments. \\

Apple &
\textasciitilde{}20 data centers (U.S., Europe) &
Mid-scale, vertically integrated &
$<$2\% (internal capacity) &
Emphasis on on-device/private AI; investment in internal ``AI Compute Fabric'' for LLM training and edge inference; limited public-cloud exposure. \\
\bottomrule
\end{tabularx}
\begin{tablenotes}\footnotesize
\item \textit{Note:} Market share values are indicative and derived from cloud-infrastructure and hyperscale-capacity data (Statista 2025; Brightlio 2024; Synergy Research 2024). Exact operator-level shares for total data-center capacity are not publicly disclosed, so these proxies reflect each firm’s relative scale within the global data-center ecosystem.These six firms account for nearly 70–75 \% of the world’s hyperscale or cloud-linked data-center capacity, which rationalizes their inclusion.According to Synergy Research Group, hyperscale operators now account for about 41 \% of worldwide data-center capacity.
Exact operator-level shares for total data-center capacity are not publicly disclosed; figures represent each firm's relative scale within the global hyperscale ecosystem. 
\end{tablenotes}
\label{tab:SI_firm_summary}
\end{threeparttable}

\break
\begin{figure}[H]
\centering
\includegraphics[width=0.8\textwidth]{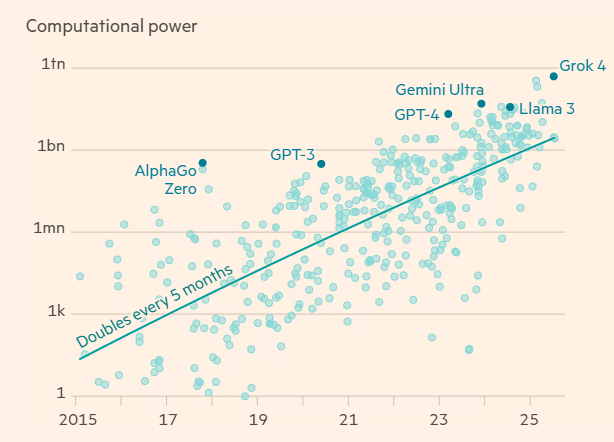}
\caption{Data includes selected AI models and is shown in log scale. Estimates are based on publicly reported numbers and calculations by Epoch.ai.     Source: Epoch.ai, adapted from \textit{Financial Times} (\url{https://ig.ft.com/ai-data-centers/}).}\label{SI_Fig2}
\end{figure}
\begin{figure}[H]
    \centering
    \includegraphics[width=0.8\textwidth]{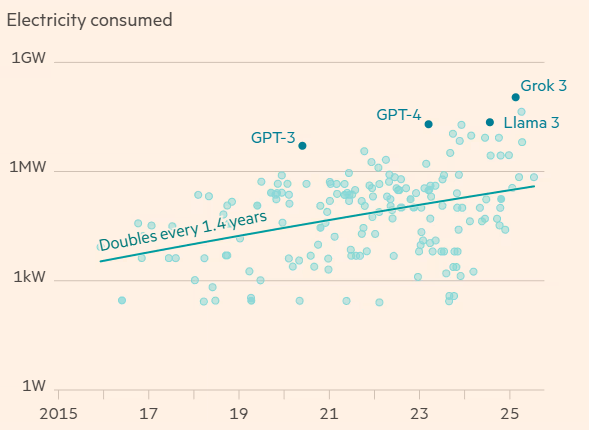}
    \caption{Computation power of leading AI systems, measured in petaflops (10\textsuperscript{15} floating-point operations per second). 
    Source: Epoch.ai, adapted from \textit{Financial Times} (\url{https://ig.ft.com/ai-data-centers/}).}
    \label{SI_Fig2}
\end{figure}

\end{document}